\documentclass{svproc}
\usepackage{url}

\usepackage{graphicx}
\usepackage{mathrsfs}
\usepackage{multirow}
\usepackage{bm}
\usepackage{amsmath}
\usepackage{amsfonts}
\usepackage{amssymb}

\newcommand{\R}{\mathbb{R}}

\begin{document}
\mainmatter              
\title{Hamilton-Jacobi-Bellman Equation Arising from Optimal Portfolio Selection Problem}
\titlerunning{HJB Equation Arising from Optimal Portfolio Selection Problem}  
%
\author{
Daniel \v{S}ev\v{c}ovi\v{c}\inst{1}
\and 
Cyril Izuchukwu Udeani\inst{1}
}
\authorrunning{Daniel \v{S}ev\v{c}ovi\v{c} and Cyril Izuchukwu Udeani} 
%
\tocauthor{Daniel \v{S}ev\v{c}ovi\v{c}, Cyril Izuchukwu Udeani}
\institute{
Comenius University in Bratislava, Mlynsk\'a dolina, 84248 Bratislava, Slovakia, \\
\email{sevcovic@fmph.uniba.sk},\\ 
\texttt{www.iam.fmph.uniba.sk/institute/sevcovic}
}

\maketitle              

\begin{abstract}
The Hamilton-Jacobi-Bellman equation arising from the optimal portfolio selection problem is studied by means of the maximal monotone operator method. The existence and uniqueness of a solution to the Cauchy problem for the nonlinear parabolic partial integral differential equation in an abstract setting are investigated by using the Banach fixed-point theorem, the Fourier transform, and the monotone operators technique.

\end{abstract}
\section{Introduction}
In this survey paper, we investigate the existence and uniqueness of a solution $\varphi=\varphi(x,\tau)$ to the Cauchy problem for the nonlinear parabolic PDE
\begin{equation}
\label{generalPDE}
\partial_{\tau}\varphi -\partial_x^2 \alpha(\varphi) = g_0(\varphi) + \partial_x g_1( \varphi),\quad 
\varphi(x, 0) =\varphi_{0}(x),
\end{equation}
where  $\tau\in(0,T), x\in\mathbb{R}$, $ g_0, g_1$ are Lipchitz continuous functions. 
The diffusion function $\alpha=\alpha(x,\varphi)$ is assumed to be Lipschitz continuous and strictly increasing in the $\varphi$-variable. In this contribution, we focus our attention to the case when the value function of the following parametric optimization problem is of the form:
\begin{equation}
\alpha(x,\varphi) = \min_{ {\bm{\theta}} \in \triangle} 
\left(
-\mu(x,{\bm{\theta}}) +  \frac{\varphi}{2}\sigma(x,{\bm{\theta}})^2\right), \quad x\in\mathbb{R} .
\label{eq_alpha_general}
\end{equation}
Here, $\mu$ and $\sigma^2$ are given $C^1$ functions, and $\triangle\subset\mathbb{R}^n$ is a compact decision set. In the case when the mean return vector $\bm{\mu}$ and the covariance matrix $\bm{\Sigma}$ respectively belong to some compact uncertainty sets $\mathscr{M}$ and $\mathscr{S}$, the value function for the worst case portfolio optimization has the form:
\[
\alpha(\varphi) = \min_{ \bm{\theta}
\in \triangle} \left(\max_{\bm{\mu}\in\mathscr{M}, \bm{\Sigma}\in\mathscr{S}}\left( -\bm{\mu}^T \bm{\theta} + \frac{\varphi}{2}\bm{\theta}^T \bm{\Sigma} \bm{\theta}\right) \right)
\]
(cf. Kilianov\'a and Trnovsk\'a \cite{KilianovaTrnovska}). The nonlinear parabolic equation expressed in (\ref{generalPDE}) is a result of dynamic stochastic programming. Assume that the underlying stochastic process $\{x_t^{{\bm{\theta}}}\}$ satisfies the following It\^{o}'s  stochastic differential equation
$d x_t^{\bm{\theta}} = \mu(x_t^{\bm{\theta}}, {\bm{\theta}}_t) dt + \sigma(x_t^{\bm{\theta}},  {\bm{\theta}}_t) dW_t$,  where the control process $\{{\bm{\theta}}_t\}$ is adapted to the process $\{x_t\}$. Here, $\mu(x,{\bm{\theta}})$ and $\sigma(x,{\bm{\theta}})$ are the drift and volatility functions, respectively, and $\{W_t\}$ is the standard one-dimensional Wiener process. We assume that the control parameter ${\bm{\theta}}$ belongs to a given compact subset $\triangle$ in $\mathbb{R}^n$. Our goal is to maximize the conditional expected value of the terminal utility of the portfolio:
\begin{equation}
\max_{{\bm{\theta}}|_{[0,T)}\subset\triangle} \mathbb{E}
\left[u(x_T^{\bm{\theta}})\, \big| \, x_0^{\bm{\theta}}=x_0 \right],
\label{maxproblem}
\end{equation}
on a finite time horizon $[0,T]$, where $u: \mathbb{R} \to \mathbb{R}$ is an increasing terminal utility function. Following Bertsekas \cite{Bertsekas}, we have that the intermediate value function $V(x,t):= \sup_{  {\bm{\theta}}|_{[t,T)}\subset\triangle} \mathbb{E}\left[u(x_T^{\bm{\theta}}) | x_t^{\bm{\theta}}=x \right]$ satisfies the fully nonlinear Hamilton-Jacobi-Bellman (HJB) parabolic equation 
\begin{equation}
\label{eq_HJB}
\partial_t V + \max_{ {\bm{\theta}} \in \triangle} 
\left(
\mu(x,t,{\bm{\theta}})\, \partial_x V 
+ \frac{1}{2} \sigma(x,t,{\bm{\theta}})^2\, \partial_x^2 V \right) = 0\,,  
\quad V(x,T)=u(x),  
\end{equation}
where $x\in\mathbb{R}, t\in [0,T)$. A typical example of the decision set $\triangle$ is the compact convex simplex $\triangle\equiv\mathcal{S}^n = \{{\bm{\theta}} \in \mathbb{R}^n\  |\  {\bm{\theta}} \ge \mathbf{0}, \mathbf{1}^{T} {\bm{\theta}} = 1\} \subset \mathbb{R}^n$, where $\mathbf{1} = (1,\dots,1)^{T} \in \mathbb{R}^n$. To solve the Cauchy problem (\ref{eq_HJB}), we can employ the Riccati transformation. Following the papers by Abe and Ishimura \cite{AI}, Ishimura and \v{S}ev\v{c}ovi\v{c} \cite{IshSev}, \v{S}ev\v{c}ovi\v{c} and Macov\'a \cite{MS}, and Kilianov\'a and \v{S}ev\v{c}ovi\v{c} \cite{KilianovaSevcovicANZIAM}, the Riccati transformation of the value function reads as follows:
\begin{equation}
\varphi(x,\tau) = - \partial_x^2 V(x,t)/\partial_x V(x,t), \quad\hbox{where}\ \ \tau=T-t.
\label{eq_varphi}
\end{equation}
According to \cite[Theorem 4.2]{KilianovaSevcovicKybernetika}, an intermediate value function $V(x,t)$ such that $\partial_x V>0$ is a solution to the Hamilton-Jacobi-Bellman equation (\ref{eq_HJB}) if and only if the transformed function $\varphi(x,\tau)$, is a solution to the Cauchy problem for the quasilinear parabolic PDE:
\begin{equation}
\label{finalequation}
\partial_\tau \varphi - \partial^2_x \alpha(\cdot,\varphi) = - \partial_x \left( \alpha(\cdot,\varphi)\varphi\right),
\quad \varphi(x,0) = \varphi_0(x),\quad (x,\tau)\in\mathbb{R}\times(0,T). 
\end{equation}
Equation (\ref{finalequation}) is of the form (\ref{generalPDE}) with $g_0$ being equal to zero and $g_1(\cdot,\varphi) = -\alpha(\cdot, \varphi)\varphi$.

\section{The value function $\alpha$ and static Markowitz model}
\begin{figure}
\centering
\includegraphics[height=1.9truecm]{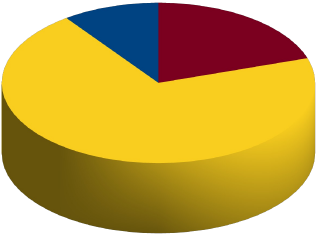}
\includegraphics[height=1.9truecm]{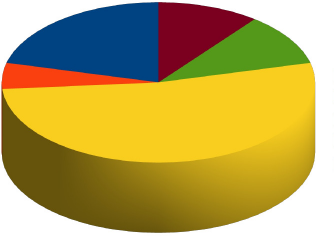}
\includegraphics[height=1.9truecm]{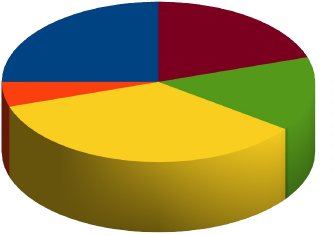}
\includegraphics[height=1.9truecm]{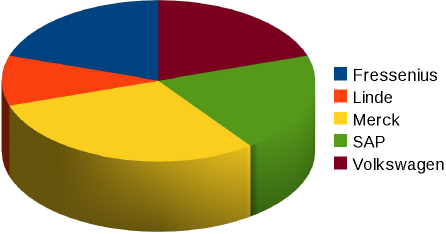}

$\varphi = 1$ \hglue1.9truecm $\varphi = 4$ \hglue 1.9truecm $\varphi = 6$ \hglue2truecm $\varphi = 8$ \hglue 1truecm \ 

\caption{Optimal asset allocation for the German DAX30 stock index for various $\varphi>0$. $\varphi$ can be viewed as risk aversion. Source: own calculations.}
\label{fig:pies}
\end{figure}

Recall that the goal of the classical Markowitz static optimization model is to maximize the mean return of the set of stochastic returns $X^i, i=1,\ldots, n$, under the constraint that the variance of the portfolio is bounded by a given constant $\sigma_0^2$. Given a vector $\bm{\theta}=(\theta_1, \ldots, \theta_n)^T $ of weights, we construct a portfolio $X =\sum_{i=1}^n \theta_i X^i$. Let $\bm{\mu}\in\mathbb{R}^n, \mu_i=\mathbb{E}(X^i)$, be the vector of mean returns of stochastic asset returns and $\bm{\Sigma}$ be their covariance matrix, $\bm{\Sigma}_{ij} = cov(X^i, X^j)$, then $\mathbb{E}(X) = \bm{\mu}^T \bm{\theta}$, and the variance $\mathbb{D}(X) = \bm{\theta}^T\bm{\Sigma} \bm{\theta}$. The Markowitz optimal portfolio optimization problem can be formulated as the following convex optimization problem to maximize the mean return under the constraint on the variance:
\[
\max_{\bm{\theta}\in\triangle} \bm{\mu}^T \bm{\theta} \qquad \hbox{s.t.}\ \frac12 \bm{\theta}^T\bm{\Sigma} \bm{\theta} \le  \frac12\sigma_0^2,
\]
where $\triangle =\{ \bm{\theta} \in\R^n,\ \sum_{i=1}^n \theta_i = 1, \theta_i\ge 0\}$. The Lagrange function for the minimization of $-\bm{\mu}^T\bm{\theta}$ has the form:
${\mathcal L}(\bm{\theta}, \varphi, \lambda, \xi) = - \bm{\mu}^T \bm{\theta} + { \varphi \frac12 \bm{\theta}^T\bm{\Sigma} \bm{\theta}} + \lambda \bm{1}^T\bm{\theta} + \xi^T \bm{\theta}$, where $\varphi\in\R, \lambda\in\mathbb{R}, \xi\in\mathbb{R}^n$, and $\xi\ge0$ are Lagrange multipliers. The same Lagrange function corresponds to the minimization problem:
\[
\alpha(\varphi) := \min_{\bm{\theta}\in \triangle } - \bm{\mu}^T \bm{\theta} + { \frac{ \varphi}{2} \bm{\theta}^T\bm{\Sigma} \bm{\theta}}\\
\]
provided the Lagrange multiplier $\varphi>0$ is given. In Fig.~\ref{fig:pies}, we present the optimal asset allocation for the German DAX30 (2017) stock index for various values of $\varphi>0$.  The value of the Lagrange multiplier $\varphi$ can be viewed as a measure of the investor's risk aversion (see Fig.~\ref{fig:pies}). Therefore, the higher the value of risk aversion, the more the portfolio is diversified among less risky assets with lower mean returns.  

The smoothness of the value function $\alpha$ depends on the structure of the decision set $\triangle$. In general, it is only $C^{0,1}$ (Lipschitz) continuous, provided that $\triangle$ is a compact decision set. Indeed, if we define $\alpha^{\bm{\theta}}(x,\varphi) :=  -\mu(x,{\bm{\theta}}) +  \frac{\varphi}{2}\sigma(x,{\bm{\theta}})^2$ then the minimal function $\alpha(x,\varphi) = \min_{ {\bm{\theta}} \in \triangle} \alpha^{\bm{\theta}}(x,\varphi)$ is Lipschitz continuous.

\begin{figure}
\centering
\includegraphics[height=3.35truecm]{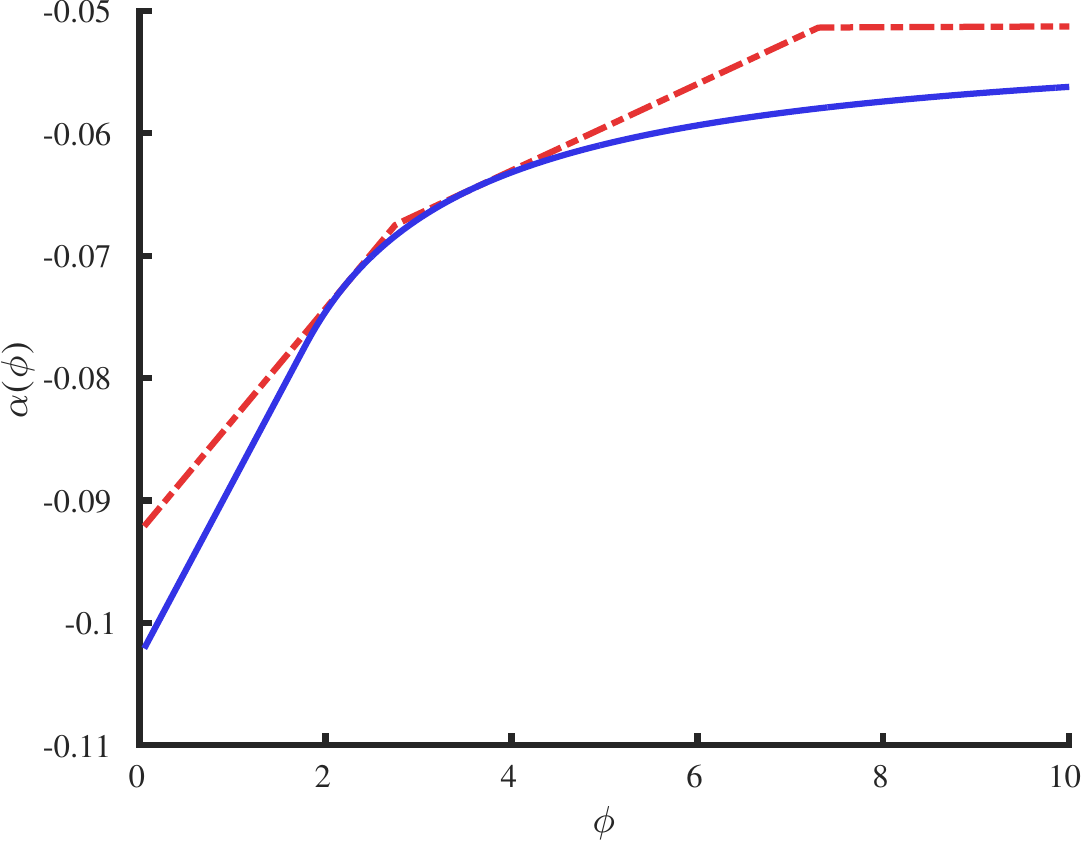}
\includegraphics[height=3.35truecm]{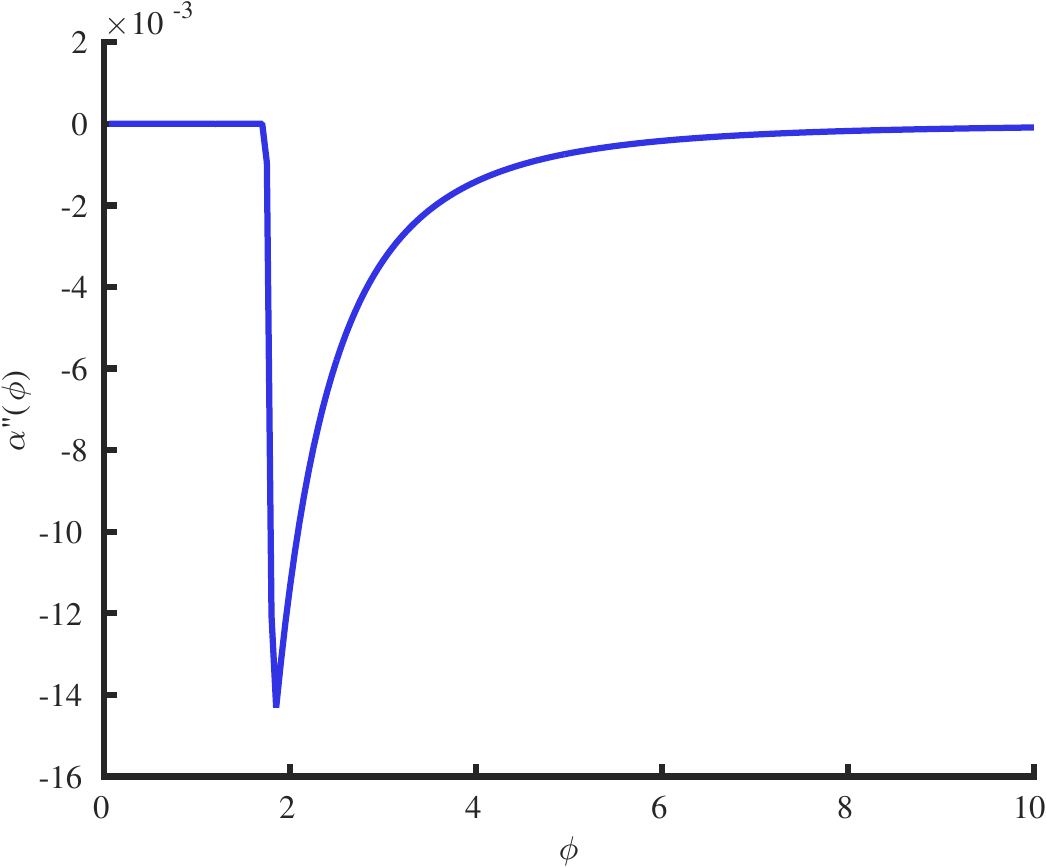}
\includegraphics[height=3truecm]{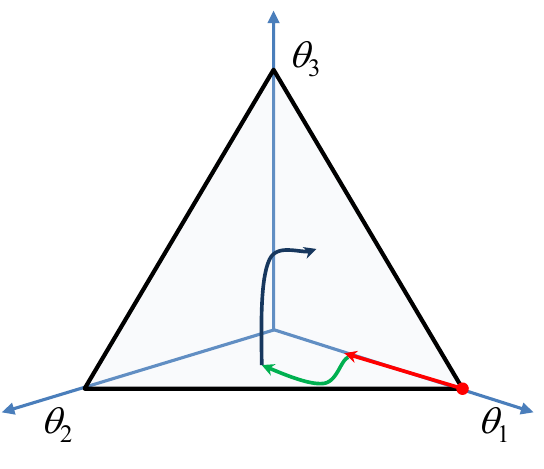}

    \small a) \hglue2.5truecm \qquad b) \hglue2.5truecm \qquad c)
    
    \caption{a) A graph of the value function $\alpha$, b) its second derivative $\alpha''(\varphi)$ for the portfolio consisting of the stocks index and bonds (cf. \cite{KMS}) for the convex compact decision set $\triangle$. The dotted line in a) corresponds to the discrete decision set $\hat\triangle=\{ {\bm{\theta}}^1, {\bm{\theta}}^2, {\bm{\theta}}^3\}\subset \triangle$. The trajectory of the minimizer $\hat{\bm{\theta}}(x,\varphi)$ for increasing $\varphi>0$ is shown in c).
     }
    \label{fig:alpha_alphader_alphaderder}
\end{figure}

In \cite[Theorem 1]{KilianovaSevcovicJIAM}, Kilianov\'a and \v{S}ev\v{c}ovi\v{c} derived sufficient conditions for the decision set $\triangle$ and functions $\mu$ and $\sigma$ that guarantee higher smoothness of the value function $\alpha$. Suppose that $\triangle\subset\mathbb{R}^n$ is a compact convex set and $\mu(x,\bm{\theta})$ and $\sigma(x,\bm{\theta})^2$ are $C^{1,1}$ smooth functions such that the function $\bm{\theta} \mapsto \mu(x,{\bm{\theta}}) + \frac{\varphi}{2} \sigma(x,{\bm{\theta}})^2$ is strictly convex. Then, the function $\alpha(\varphi)$ is $C^{1,1}$ continuous. The proof is based on the classical envelope theorem due to Milgrom and Segal \cite{milgrom_segal2002} and the result on Lipschitz continuity of the minimizer $\hat{\bm{\theta}}(x,\varphi)$ belonging to a convex compact set $\triangle$ due to Klatte \cite{Klatte}. In Fig.~\ref{fig:alpha_alphader_alphaderder}, we plot the value function $\alpha$ for convex and discrete decision sets $\triangle$, its second derivative $\alpha''$, and the dependence of the optimal decision vector $\hat{\bm{\theta}}$ on the parameter $\varphi$. If the decision set $\triangle$ is a compact convex set, then the function $\alpha(\varphi)$ is $C^{1,1}$ continuous, i.e., the derivative $\alpha^\prime(\varphi)$ is Lipschitz continuous (blue line). If $\triangle$ is a discrete subset  $\hat\triangle=\{ {\bm{\theta}}^1, {\bm{\theta}}^2, {\bm{\theta}}^3\}\subset \triangle$, then the function $\alpha$ is just $C^{0,1}$ continuous piece-wise affine function (dotted line). Furthermore, we show a trajectory of the minimizer $\hat{\bm{\theta}}(\varphi)$ for increasing $\varphi>0$. Fig.~\ref{fig:alpha_alphader_alphaderder}(c) demonstrates that for small values of $\varphi$, the minimizer belongs to a one-dimensional set (edge). For higher values of $\varphi$, the minimizer $\hat{\bm{\theta}}(\varphi)$ belongs to higher-dimensional subsets (face, volume) of the simplex $\triangle = \{{\bm{\theta}} \in \mathbb{R}^n\  |\  {\bm{\theta}} \ge \mathbf{0}, \mathbf{1}^{T} {\bm{\theta}} \le  1\}$.

\section{Existence and uniqueness of solutions in Sobolev spaces}
Let $V\hookrightarrow H \hookrightarrow V^\prime$ be the so-called Gelfand triple, where
$H = L^2(\mathbb{R}) = \{ f:\mathbb{R}\to\mathbb{R}, \Vert f\Vert_{L^2}^2 = \int_{\mathbb{R}} |f(x)|^2 dx < \infty \}$ is a Hilbert space endowed with the inner product $(f,g)= \int_{\mathbb{R}} f(x) g(x) dx$. Here, $V= H^1(\mathbb{R})$ is a Sobolev space, and $V^\prime= H^{-1}(\mathbb{R})$ is its dual space. The triple $V\hookrightarrow H \hookrightarrow V^\prime$ naturally induces the time-dependent Gelfand triple $\mathcal{V}\hookrightarrow \mathcal{H} \hookrightarrow \mathcal{V^\prime}$, where $ \mathcal{H}$ is a Hilbert space endowed with the norm $\Vert \varphi\Vert^{2}_{\mathcal{H}} = \int_{0}^{T}\Vert \varphi(\tau)\Vert^{2}_{H}d\tau, \; \forall\varphi\in\mathcal{H}$. Similarly, we define the spaces $\mathcal{V} = L^2 ((0,T);V)$, $\mathcal{H} = L^{2}((0,T);H) $ and $\mathcal{V^\prime} = L^2 ((0,T);V^\prime)$. 

In \cite{udeani2021application}, Udeani and \v{S}ev\v{c}ovi\v{c} proved the following result on the existence and uniqueness of a solution to (\ref{finalequation}). The proof is based on the maximal monotone operator technique (cf. Barbu \cite{Barbu} and Showalter \cite{Showalter}) and the Banach fixed-point theorem. 

Let us introduce the auxiliary functions: $p(x)= \max_{{\bm{\theta}}\in\triangle} |\partial_x\mu(x,{\bm{\theta}})|$, and $h(x)=  -\max_{{\bm{\theta}}\in\triangle} \mu(x,{\bm{\theta}})$.

\begin{theorem}\cite[Theorem 5]{udeani2021application}
Let the decision set $\triangle \subset \mathbb{R}^n$ be compact and the function $u:\mathbb{R}\to\mathbb{R}$ be an increasing utility function such that $\varphi_0(x) = -u''(x)/u'(x)$ belongs to the space $L^2(\mathbb{R})\cap L^\infty(\mathbb{R})$. Suppose that the drift $\mu(x, \bm{\theta})$ and volatility function $\sigma^2(\bm{\theta})>0$ are $C^1$ continuous in the $x$ and $\bm{\theta}$ variables, and the value function $\alpha(x,\varphi)$ given in (\ref{eq_alpha_general}) satisfies $p\in L^2(\mathbb{R})\cap L^\infty(\mathbb{R}), h\in L^\infty(\mathbb{R}), \; \text{and} \;\partial^2_x h \in L^2(\mathbb{R})$. 
Then for any $T>0$ there exists a unique solution $\varphi$ of the Cauchy problem (\ref{finalequation}) that satisfies $\varphi\in C([0,T]; H)\cap L^2((0,T); V)\cap L^\infty((0,T)\times \mathbb{R})$.
\end{theorem}

\begin{figure}
    \centering

    \includegraphics[width=0.45\textwidth]{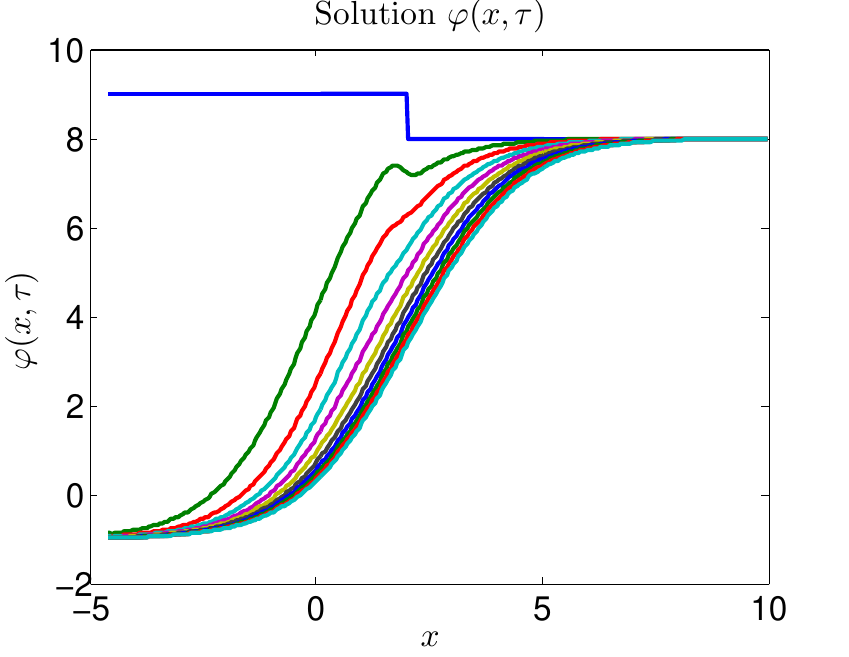}       
    \caption{A solution $\varphi(x,\tau)$ for the DARA utility function
    $u$ such that $\varphi_0(x)=-u''(x)/u'(x)\in \{9,8\}$. Source: our calculations based on the numerical method from \cite{udeani2021application,KilianovaSevcovicANZIAM}}
    \label{fig:vysledky-DAX}
\end{figure}

In contrast to the fully nonlinear property of the original HJB equation (\ref{eq_HJB}), the transformed equation (\ref{finalequation}) represents a quasilinear parabolic equation in divergence form. Thus, efficient numerical schemes can be constructed for this class of equations. In our computational experiments, we employ the finite volume discretization scheme proposed and investigated by Kilianov\'a and \v{S}ev\v{c}ovi\v{c} \cite{KilianovaSevcovicKybernetika,KilianovaSevcovicANZIAM,KilianovaSevcovicJIAM}). Fig.~\ref{fig:vysledky-DAX} shows the results of a time-dependent sequence of profiles $\varphi(x,\tau)$ for a constant initial condition $\varphi_0\equiv 9$. This graph shows the solution profiles for the discountinuous initial condition $\varphi_0\in \{9,8\}$. It represents the utility function  $u$  of the decreasing absolute risk aversion (DARA) such that $\varphi_0(x)=-u''(x)/u'(x)$. The function $\varphi(x,\tau)$ increases in the variable $x$ and decreases in the variable $\tau=T-t$. Therefore, the optimal vector $\bm{\theta}(x,\tau)$ contains a more diversified portfolio of assets when $x$ increases and time $t\to T$ (see Fig.~\ref{fig:vysledky-DAX}). Furthermore, it is reasonable to invest in an asset with the highest expected return when the value of the account $x$ is low, while an investor must diversify the portfolio when $x$ is large and time $t$ approaches the end of maturity $T$.

\section{Conclusions}
In this paper, we discussed the qualitative and numerical results of a fully nonlinear HJB equation that arises from a stochastic dynamic optimization problem in Sobolev spaces. This equation is related to a portfolio management problem where the goal is to maximize the expected terminal utility of a portfolio. We transformed the equation into a quasilinear parabolic equation using the Riccati method, and under certain assumptions, we showed that the diffusion function is globally Lipschitz continuous. We also provided numerical examples to illustrate our results.

\medskip
\noindent{\bf Acknowledgments.} Support from the Slovak Research and Development Agency under the project APVV-20-0311 (C.U.) and the VEGA 1/0611/21 grant (D.\v{S}.) are kindly acknowledged.

%

\end{document}